\documentclass{elsart}
\usepackage{epsfig}
\usepackage{amssymb}
\journal{Physics Letters B}

\def\ac{$\alpha$+$^{12}$C\ }
\def\ent{$\alpha$+$^{12}$C$_{\rm g.s.}$\ }
\def\ext{$\alpha$+$^{12}$C$^*(0^+_2)$\ }
\def\2ext{$\alpha$+$^{12}$C$^*(2^+_1)$\ }
\def\AA{nucleus-nucleus\ }
\def\aA{$\alpha$-nucleus\ }

\def\oo{$^{16}$O+$^{16}$O\ }
\def\o17o15{$^{16}$O($^{16}$O,$^{17}$O)$^{15}$O$^*$\ }

\begin{document}
\begin{frontmatter}
\vspace*{-1cm}
\title{Missing monopole strength of the Hoyle state in the inelastic
\ac scattering\thanksref{label1}}
\thanks[label1]{Research supported, in part, by Natural Science Council
of Vietnam, EU Asia-Link Program CN/ASIA-LINK/008 (94791) and Vietnam Atomic
Energy Commission (VAEC).}

\author{Dao T. Khoa\corauthref{cor1}}
\ead{khoa@vaec.gov.vn}\corauth[cor1]{Corresponding author,}
 and \author{Do Cong Cuong}
\address{Institute for Nuclear Science and Technique, VAEC,
P.O. Box 5T-160, Nghia Do, Hanoi, Vietnam.}

\begin{abstract} Analyses of the inelastic \ac scattering at medium
energies have indicated that the strength of the Hoyle state (the isoscalar
0$^+_2$ excitation at 7.65 MeV in $^{12}$C) seems to exhaust only 7 to 9\% of
the monopole energy weighted sum rule (EWSR), compared to about 15\% of the EWSR
extracted from inelastic electron scattering data. The full monopole transition
strength predicted by realistic microscopic $\alpha$-cluster models of the Hoyle
state can be shown to exhaust up to 22\% of the EWSR. To explore the missing
monopole strength in the inelastic \ac scattering, we have performed a fully
microscopic folding model analysis of the inelastic \ac scattering at $E_{\rm
lab}=104$ to 240 MeV using the 3-$\alpha$ resonating group wave function of the
Hoyle state obtained by Kamimura, and a complex density-dependent M3Y
interaction newly parametrized based on the Brueckner Hartree Fock results for
nuclear matter. Our folding model analysis has shown consistently that the
missing monopole strength of the Hoyle state is not associated with the
uncertainties in the analysis of the \ac scattering, but is most likely due to
the short lifetime and weakly bound structure of this state which significantly
enhances absorption in the exit \ext channel.
\end{abstract}
\end{frontmatter}

Given a vital role in the stellar synthesis of Carbon, the isoscalar 0$^+_2$
state at 7.65 MeV in $^{12}$C (known as the Hoyle state) has been studied over
the years in numerous experiments. Although this state was clearly identified
long ago in the inelastic \ac scattering at medium energies
\cite{Hau69,Spe71,Smi73,Wik81} and inelastic electron scattering \cite{Stre70}
as an isoscalar $E0$ excitation, our knowledge about its unique structure is
still far from complete \cite{Fre07}. Since the Hoyle state lies slightly above
the $\alpha$-decay threshold of 7.27 MeV, its wave function should have a
dominant $\alpha$ cluster component. In fact, the Hoyle state has been well
described in terms of three $\alpha$ clusters by the Resonating Group Method
(RGM) some thirty years ago \cite{Ueg77,Kam81}. Very interesting is the
condensate scenario suggested recently \cite{Toh01,Fun03} where the three
$\alpha$ clusters were shown to condense into the lowest ($s$-state) of their
potential, and thus, forming a Bose-Einstein condensate (BEC). A more
complicated structure of the Hoyle state was shown by the Fermionic Molecular
Dynamics (FMD) calculation \cite{Che07} where the BEC wave function is mixed
also with the molecular $^8$Be$+\alpha$ configuration. In general, to validate
conclusion made in the structure calculation, the wave functions must be
carefully tested in the study of nuclear reactions exciting the Hoyle state. In
this aspect, the model wave functions of the Hoyle state
\cite{Kam81,Che07,Fun06} have been shown to give a reasonable description of the
inelastic electron scattering data. The electric monopole transition moment
\begin{equation}
 M(E0,0^+_1\to 0^+_2)=e\sqrt{4\pi}
 \int_0^\infty{\rho^{\rm proton}_{0^+_1\to 0^+_2}(r)r^4dr}
 \label{e0}
\end{equation}
predicted by the RGM \cite{Kam81}, BEC \cite{Fun03} and FDM \cite{Che07}
calculations is around 6.62, 6.45 and 6.53 $e$~fm$^2$, respectively, which are
about 20\% larger\footnote{a direct comparison of (\ref{e0}) with the observed
$M(E0)$ can be made if the small difference between the charge- and proton
transition densities is neglected.} than the experimental moment $M(E0)\approx
5.37\pm 0.22\ e$~fm$^2$ deduced from the $(e,e')$ data \cite{Stre70}. The
inelastic electron scattering probes, however, only the charge transition
density and it is necessary to study also inelastic hadron scattering which
probes the nuclear transition density. Such experiments have been done, e.g.,
for inelastic $^{3,4}$He,$^6$Li+$^{12}$C scattering \cite{Leb80,Eyr87,John03}
and the data analyses, either in the distorted wave Born approximation (DWBA) or
coupled-channel (CC) formalism, were performed using the collective model form
factor (FF) for the Hoyle state. These analyses indicated that the observed
monopole transition strength of the Hoyle state exhausts about 7 to 9\% of the
isoscalar monopole energy weighted sum rule (EWSR). In particular, the DWBA
analysis of the inelastic \ac data, measured recently at $E_{\rm lab}=240$ MeV
with high precision \cite{John03}, has found that the Hoyle state exhausts
$7.6\pm 0.9\%$ of the EWSR. We recall that the sum rule fraction $S_0$ of a
monopole excitation is determined \cite{Sat83} as
\begin{equation}
 S_0=E_{\rm x}|M(IS0,0^+_1\to 0^+_2)|^2/\left(\frac{2\hbar^2A<r^2>}{m}\right),
 \label{SumR}
\end{equation}
where $E_{\rm x}, m$ and $A$ are the excitation energy, nucleon mass and target
mass number, respectively, $<r^2>$ is the mean square radius and $M(IS0,0^+_1\to
0^+_2)$ is the isoscalar monopole transition moment which is determined by the
same Eq.~(\ref{e0}) but using the nuclear transition density instead of the
charge transition density. If we assume $M(IS0,0^+_1\to 0^+_2)\approx
2M(E0,0^+_1\to 0^+_2)/e$ and take $<r^2>\approx 5.78$ fm$^2$ (estimated using
the ground state density of $^{12}$C given by the RGM \cite{Kam81}), then we
obtain $S_0\approx 22.8, 21.7$ and 22.2\% of the EWSR from the RGM \cite{Kam81},
BEC \cite{Fun03} and FDM \cite{Che07} results, respectively, and $S_0\approx
15.0\pm 1.3\%$ of the EWSR from the experimental monopole moment, in a perfect
agreement with $S_0$ value given in Ref.~\cite{Stre70}. Given a good description
of the $(e,e')$ data by the cluster models \cite{Kam81,Toh01,Fun03,Che07}, we
conclude that the monopole strength of the Hoyle state should exhaust 15 to 20\%
of the EWSR and a puzzle remains why at least half of the monopole strength of
the Hoyle state is missing in the inelastic \ac scattering.

To investigate the missing monopole strength of the Hoyle state, we have
performed a consistent folding model analysis of the inelastic \ac scattering at
$E_{\rm lab}=104$ to 240 MeV using microscopic nuclear densities obtained from
the 3-$\alpha$ RGM wave functions by Kamimura \cite{Kam81}, and a new complex
density-dependent M3Y interaction. Before discussing our results, we recall that
the same RGM nuclear densities \cite{Kam81} were used in a number of folding
model studies (see, e.g., Refs.~\cite{Bau84,Hir02,Oh04,Tak06,Oh07}) of the
inelastic \ac scattering. Excepting the early work by Bauhoff \cite{Bau84} where
a density independent Gaussian has been used as $\alpha N$ interaction in the
single-folding calculation, other studies mentioned here have used the
well-known DDM3Y \cite{Ko82}  interaction in the double-folding calculation of
the \ac potentials. Since the DDM3Y interaction is real, the imaginary parts of
both the optical potential (OP) and inelastic FF were chosen phenomenologically
in these studies. For example, in the CC analysis of the \ac scattering by
Ohkubo and Hirabayashi \cite{Hir02,Oh04} the imaginary inelastic FF was
neglected and parameters of the Woods-Saxon imaginary OP were adjusted
separately for each exit channel to obtain a good CC fit to the measured cross
sections. Although one could achieve a reasonable description of the inelastic
\ac scattering data in such a CC analysis, an arbitrary choice of the imaginary
potentials makes it difficult to estimate accurately the absolute $E0$
transition strength. Up to now, the $E0$ strength of the Hoyle state has been
deduced from the DWBA or CC analyses of the inelastic $^{3,4}$He,$^6$Li+$^{12}$C
scattering \cite{Leb80,Eyr87,John03} using the breathing mode (BM) model
\cite{Sat83} for the nuclear transition density
\begin{equation}
 \rho_{0^+_1\to 0^+_2}(r)=-\beta~[3\rho_0(r)+r\frac{d\rho_0(r)}{dr}],
 \label{e1}
\end{equation}
where $\rho_0(r)$ is the ground-state (g.s.) density of $^{12}$C. In this case,
$S_0=[\beta/\beta_{\rm max}]^2$, where $\beta_{\rm max}$ is the deformation
parameter required for a monopole excitation to exhaust 100\% of the EWSR (see
Eq.~(3.5) in Ref.~\cite{Ho95}). The inelastic scattering FF is then obtained by
(double) folding the transition density (\ref{e1}) with the effective
nucleon-nucleon ($NN$) interaction and projectile g.s. density
\cite{Kho97,Kho00}. In the \aA scattering, one also uses a simpler approach to
(single) fold the density (\ref{e1}) with an appropriate $\alpha$-nucleon
($\alpha N$) interaction \cite{Sat97}. In general, the (complex) strength of the
$\alpha N$ or $NN$ interaction is first adjusted to the best optical model or CC
description of elastic scattering and then is used without any further
renormalization to calculate the inelastic FF. As a result, the only parameter
in the inelastic channel is the deformation parameter $\beta$ which must be
determined from the best DWBA or CC fit to the inelastic scattering data. Since
the dilute $\alpha$-clustered structure of the Hoyle state is different from a
monopole compression mode, the use of the BM density (\ref{e1}) in the folding
model calculation might be questionable. In addition, the uncertainties of the
effective $NN$ or $\alpha N$ interaction and the optical potential (OP) in the
entrance and exit channels could also cause some deviation in the description of
inelastic cross section which results on the missing $E0$ strength.

\begin{figure}[ht]
\centering\vspace*{2.5cm}%\hspace*{-2cm}
 \mbox{\epsfig{file=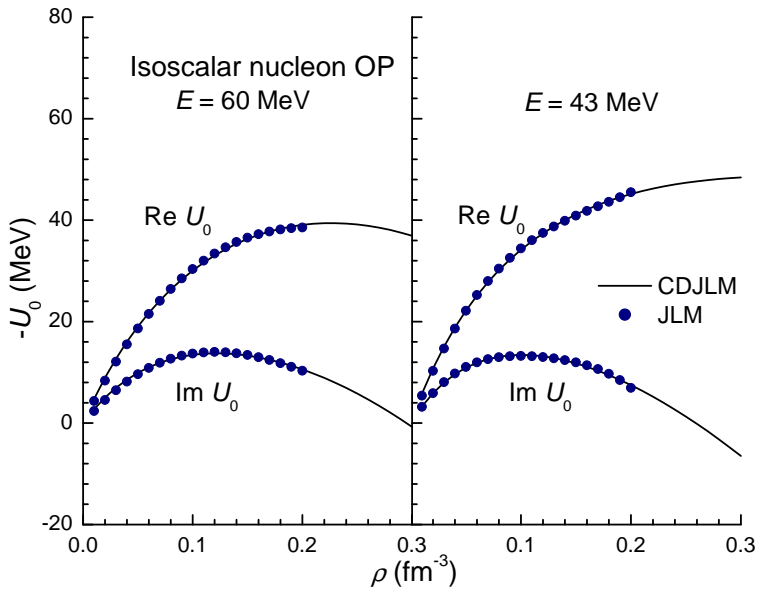,height=9cm}}\vspace*{-3cm}
\caption{Isoscalar optical potential $U_0(E,\rho)$ of nucleon incident on
nuclear matter at 43 and 60 MeV given by HF calculation (\ref{e2}) using the new
CDJLM interaction (solid curves). The points are the microscopic BHF results
made available up to nuclear matter density $\rho=0.2$ fm$^{-3}$ by the JLM
group \cite{Je77}. This CDJLM interaction has been used in the present folding
model study of the \ac scattering at $E_{\rm lab}=172.5$ and 240 MeV.}
\label{f1}
\end{figure}

We note that the RGM wave functions by Kamimura \cite{Kam81} were proven to give
consistently a realistic description of the shell-like structure of the ground
state $0^+_1$, first excited $2^+_1$ (4.44 MeV) and $3^-_1$ (9.64 MeV) states,
and the $\alpha$-clustered structure of the $0^+_2$ (7.65 MeV) Hoyle state. For
the latter, the RGM wave function has been shown \cite{Fun03,Fun06} to be very
close to the BEC wave function and both of them give, in fact, nearly the same
description of the inelastic \ac scattering \cite{Tak06}. Thus, the RGM nuclear
densities is a very good choice for the present folding model study. Concerning
the effective $NN$ interaction, a density dependent version of the M3Y-Paris
interaction (dubbed as CDM3Y6 interaction \cite{Kho97}) is the most often used
in our folding calculations. The CDM3Y6 density dependent parameters were
carefully adjusted in the Hartree-Fock (HF) calculation to reproduce the
saturation properties of nuclear matter \cite{Kho97}. However, like the DDM3Y
version, the CDM3Y6 interaction is \emph{real} and can be used to predict the
real OP and inelastic FF only. To avoid a phenomenological choice of the
imaginary potentials like those used in earlier folding model studies
\cite{Hir02,Oh04,Tak06,Oh07} of the inelastic \ac scattering, we have
constructed in the present work a new \emph{complex}, density dependent
M3Y-Paris interaction. The parameters of the complex density dependence were
calibrated against the Brueckner Hartree-Fock (BHF) results \cite{Je77} for the
nucleon OP in nuclear matter by Jeukenne, Lejeune and Mahaux (JLM) at each
considered energy. Namely, the \emph{isoscalar} complex nucleon OP in nuclear
matter is determined from the HF matrix elements of the new density dependent
$NN$ interaction as
\begin{equation}
  U_0(E,\rho)=\sum_{j\le k_F}[<kj|u_{\rm D}(E,\rho)|kj>+
  <kj|u_{\rm EX}(E,\rho)|jk>]. \label{e2}
\end{equation}
Here $k_F=[1.5\pi^2\rho]^{1/3}$ and $k$ is the momentum of the incident nucleon
which must be determined self-consistently from $U_0$ as
 $k=\sqrt{2m[E-{\rm Re}~U_0(E,\rho)]/\hbar^2}$. We have used in Eq.~(\ref{e2})
two different CDM3Y functionals \cite{Kho97} to construct separately the
\emph{real} and \emph{imaginary} parts of the density dependence, so that the
real and imaginary parts of the effective $NN$ interaction $u_{\rm D(EX)}$ are
determined as
\begin{equation}
 {\rm Re}[u_{\rm D(EX)}]=F_{\rm V}(E,\rho)v_{\rm D(EX)}(s),\
 {\rm Im}[u_{\rm D(EX)}]=F_{\rm W}(E,\rho)v_{\rm D(EX)}(s).
 \label{e4}
\end{equation}
The radial direct and exchange interactions $v_{\rm D(EX)}(s)$ were kept
unchanged, as derived from the M3Y-Paris interaction \cite{An83}, in terms of
three Yukawas (see, e.g., Ref.~\cite{Kho00,Kho07}). The parameters of the
\emph{complex} density dependence $F_{\rm V(W)}(E,\rho)$ were adjusted
iteratively until $U_0(E,\rho)$ agrees closely with the tabulated JLM results at
each energy \cite{Je77} as shown in Fig.~\ref{f1}.
More details on the new density dependent interaction, dubbed hereafter as CDJLM
interaction, and the explicit parameters of $F_{\rm V(W)}(E,\rho)$ at different
energies will be presented elsewhere. The CDJLM interaction was used to
calculate the complex OP and inelastic FF for the microscopic DWBA and CC
analyses of the elastic and inelastic \ac scattering data at $E_{\rm lab}=104$
\cite{Hau69,Spe71}, 139 \cite{Smi73}, 172.5 \cite{Wik81}, and 240 MeV
\cite{John03}. The generalized double-folding method \cite{Kho00} was used to
calculate the complex \aA potential as the following HF-type matrix element of
the CDJLM interaction (\ref{e4})
\begin{equation}
 U_{A\to A^*}=\sum_{i\in \alpha;j\in A,j'\in A^*}[<ij'|u_{\rm D}|ij>
 +<ij'|u_{\rm EX}|ji>],
\label{e5}
\end{equation}
where $A$ and $A^*$ are states of the target in the entrance- and exit channel
of the \aA scattering, respectively. Thus, Eq.~(\ref{e5}) gives the elastic
(diagonal) OP if $A^*=A$ and inelastic (transition) FF if otherwise. A more
accurate local density approximation suggested in Ref.~\cite{Kho01} has been
used for the exchange term in Eq.~(\ref{e5}). The dynamic change in the density
dependence $F_{\rm V(W)}(\rho)$ caused by the inelastic \ac scattering
\cite{Kho00,Sri84} is taken into account properly. All the DWBA and CC
calculations have been performed using the code ECIS97 written by Raynal
\cite{Raynal}.

\begin{figure}[ht]
\centering\vspace*{3cm}%\hspace*{-2cm}
 \mbox{\epsfig{file=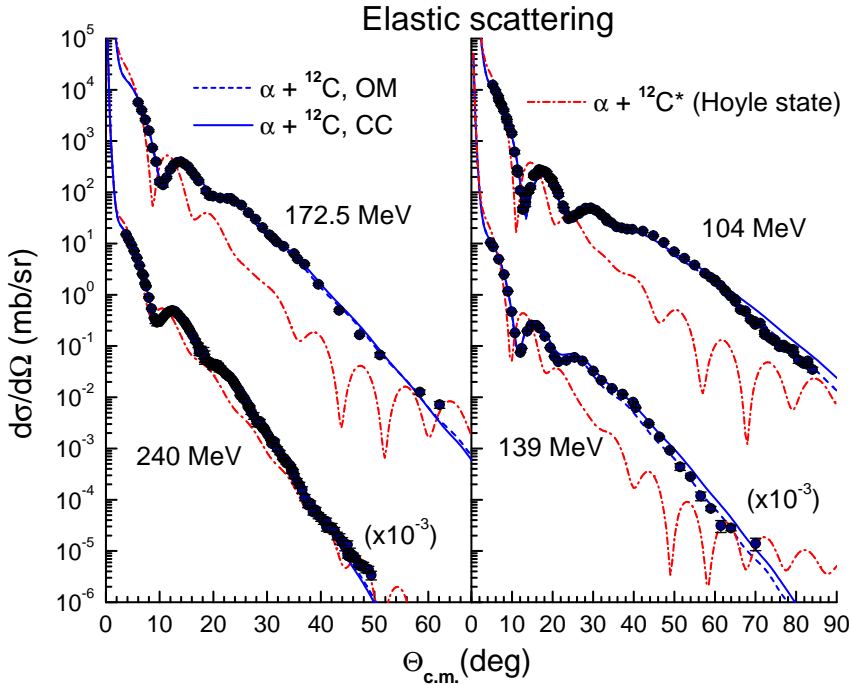,height=10cm}}\vspace*{-3.5cm}
\caption{Elastic \ac scattering data measured at $E_{\rm lab}=104$
\cite{Hau69,Spe71}, 139 \cite{Smi73}, 172.5 \cite{Wik81} and 240 MeV
\cite{John03} in comparison with the OM and CC results given by the complex
folded OP. The elastic $\alpha$ scattering on the Hoyle state is predicted by
the OM calculation using the diagonal \ext folded OP with $N_{\rm I}$ chosen
(see Table~\ref{t2}) to reproduce the measured inelastic \ext scattering cross
section in the CC calculation.} \label{f2}
\end{figure}
To fine tune the strength of complex CDJLM interaction (\ref{e4}) which has been
obtained in the nuclear matter limit (\ref{e2}), the optical model (OM) analysis
of the measured elastic \ac scattering data was done at each energy using the
complex OP determined from the elastic folded potential (\ref{e5}) as
\begin{equation}
 U_0(R)=N_{\rm R}{\rm Re}[U_{0^+_1\to 0^+_1}(R)]+
 iN_{\rm I}{\rm Im}[U_{0^+_1\to 0^+_1}(R)]+V_{\rm C}(R),
\label{e6}
\end{equation}
where $V_{\rm C}(R)$ is the elastic Coulomb potential taken, for simplicity, as
that between a point charge and a uniform charge distribution of radius $R_{\rm
C} = 1.3\times(4^{1/3}+ 12^{1/3})$ fm. The renormalization coefficients $N_{\rm
R}$ and $N_{\rm I}$ of the real and imaginary elastic folded potentials were
adjusted first to the best OM fit of the elastic scattering data and then used
further to scale the real and imaginary inelastic folded FF for the DWBA
calculation, a standard method used sofar in the folding + DWBA analyses of
inelastic \aA scattering \cite{John03,Bau84,Kho00,Sat97}. Since $N_{\rm R}$ and
$N_{\rm I}$ are an approximate way to take into account the higher-order
(dynamic polarization) contributions to the microscopic OP \cite{Kho00,Fe92},
they need to be readjusted again (to fit the elastic data) in the CC
calculation, to account for the remaining nonelastic channels which were not
included into the CC scheme. For consistency, these new $N_{\rm R}$ and $N_{\rm
I}$ factors are also used to scale the complex inelastic folded FF for the CC
calculation. The OM and CC descriptions of the elastic \ac scattering at the
considered energies are shown in Fig.~\ref{f2} and the corresponding OP
parameters are given in Table~\ref{t2}. One can see from Table~\ref{t2} that the
$N_{\rm R}$ coefficient given by the best OM fit of the elastic scattering data
is slightly above unity which indicates that the new CDJLM interaction is quite
a realistic choice for the double-folding calculation of the \aA potential. The
best fit $N_{\rm I}$ of about 1.3 to 1.5 given by the OM calculation is not
unexpected because the imaginary part of the CDJLM interaction is based on the
BHF results for nuclear matter and gives, therefore, only a ``volume"
absorption. As a result, the imaginary folded OP cannot properly account for the
surface absorption caused by inelastic scattering to the low-lying collective
excitations and transfer reactions, and the OM fit to the elastic data naturally
requires an enhanced $N_{\rm I}$ coefficient. The best-fit folded optical
potentials give volume integrals $J_{\rm R(I)}$ and total reaction cross
sections $\sigma_{\rm R}$ very close to those obtained earlier in the folding
model analysis using a phenomenological imaginary OP \cite{Kho01} or in the
model-independent Fourier-Bessel analyses of the elastic \ac scattering data
\cite{Ba89}. Thus, the elastic distorted waves given by the CDJLM folded OP
should be quite accurate for the DWBA analysis of the inelastic \ac scattering.

\begin{figure}[ht]
\centering\vspace*{3cm}%\hspace*{-2cm}
 \mbox{\epsfig{file=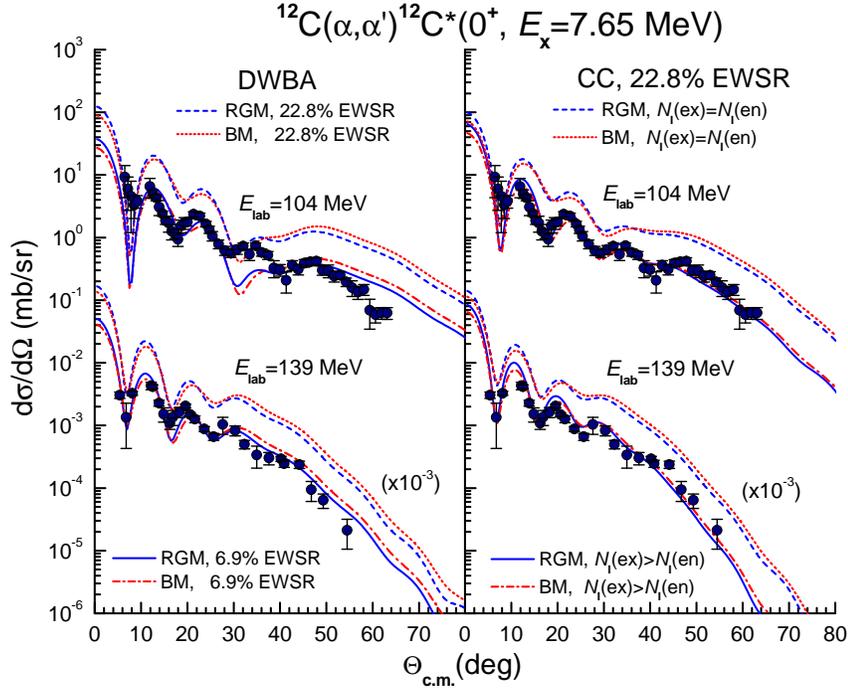,height=10cm}}\vspace*{-3.5cm}
\caption{Inelastic \ac scattering data at $E_{\rm lab}=104$ \cite{Hau69,Spe71}
and 139 \cite{Smi73} MeV for the 0$^+_2$ state of $^{12}$C in comparison with
the DWBA and CC results given by the complex folded OP and inelastic FF obtained
with two choices of the transition density for the 0$^+_2$ state. A good DWBA
description of the data is reached only if the the transition densities are
scaled by a factor of 0.55 which reduces $E0$ sum rule strength to $S_0\approx
6.9\%$ of the monopole EWSR. If $S_0$ is kept at 22.8\% of the EWSR, as given by
the RGM result \cite{Kam81}, the data can be reproduced in the CC calculation
only by using a more absorbing OP for the \ext channel. See more details in
text.} \label{f3}
\end{figure}

\begin{figure}[ht]
\centering\vspace*{3cm}%\hspace*{-2cm}
 \mbox{\epsfig{file=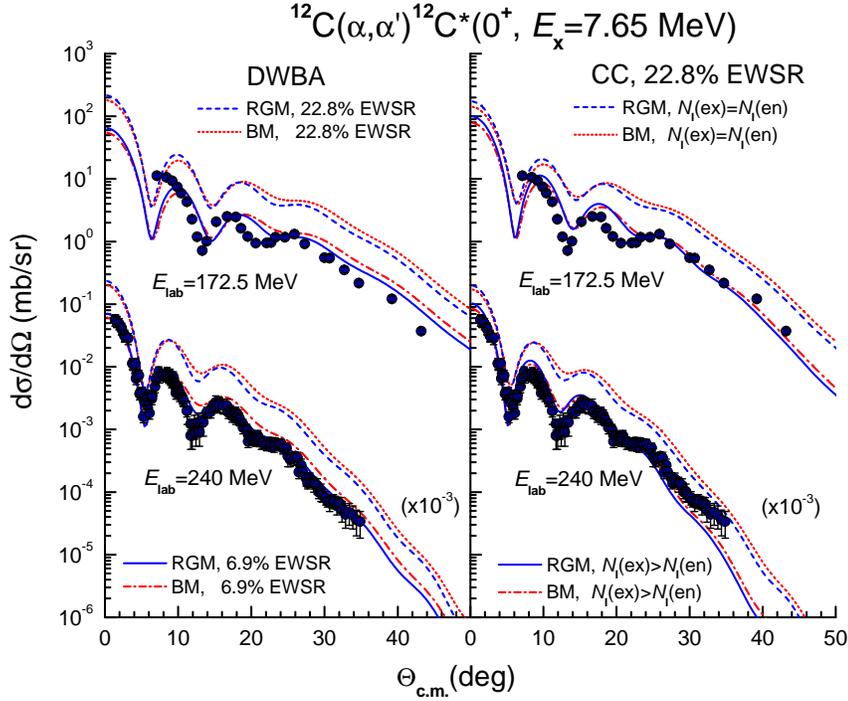,height=10cm}}\vspace*{-3.5cm}
\caption{The same as Fig.~\ref{f3} but for the data measured at $E_{\rm
lab}=172.5$ \cite{Wik81} and 240 \cite{John03} MeV.}\label{f4}
\end{figure}

\begin{table}\centering
\caption{Renormalization coefficients $N_{\rm R(I)}$ of the complex folded OP
and inelastic FF used in the DWBA and CC analyses of \ac scattering at $E_{\rm
lab}=104$, 139, 172.5, and 240 MeV. $J_{\rm R(I)}$ are volume integrals of the
real and imaginary OP, and $\sigma_{\rm R}$ is the corresponding total reaction
cross section. CC$_{\rm en}$ are the OP parameters in the entrance \ent channel
of the CC calculation. CC$_{\rm ex}$ are those of the OP in the exit \ext
channel, where $N_{\rm I}$ was adjusted to reproduce the inelastic 0$^+_2$ data
using the original RGM transition density \cite{Kam81}.}
\label{t2}\vspace*{0.3cm}
\begin{tabular}{|c|c|c|c|c|c|c|} \hline
$E_{\rm lab}$ (MeV) & Analysis & $N_{\rm R}$ & $J_{\rm R}$ (MeV~fm$^3$) &
$N_{\rm I}$ & $J_{\rm I}$ (MeV~fm$^3$) & $\sigma_{\rm R}$ (mb) \\ \hline
 104 & DWBA & 0.975 & 312.3 & 1.295 & 104.3 & 799.5 \\
    & CC$_{\rm en}$ & 1.015 & 325.1 & 1.066 & 81.76 & 794.1 \\
    & CC$_{\rm ex}$  & 1.015 & 325.1 & 2.500 & 201.4 & 1349 \\
 139 & DWBA & 1.025 & 287.4 & 1.374 & 120.2 & 770.8 \\
    & CC$_{\rm en}$ & 1.049 & 294.2 & 1.064  & 93.06 & 739.7 \\
    & CC$_{\rm ex}$  & 1.049 & 294.2 & 2.500 & 218.6 & 1287 \\
 172.5 & DWBA & 1.156 & 285.4 & 1.506 & 117.5 & 736.7 \\
    & CC$_{\rm en}$ & 1.165 & 287.6 & 1.264  & 98.64 & 716.9 \\
    & CC$_{\rm ex}$  & 1.165 & 287.6 & 3.000 & 234.1 & 1288 \\
 240 & DWBA & 1.127 & 252.8 & 1.340 & 111.0 & 659.1 \\
    & CC$_{\rm en}$ & 1.145 & 256.9 & 1.241  & 102.8 & 656.4 \\
    & CC$_{\rm ex}$  & 1.145 & 256.9 & 3.400 & 281.6 & 899.1 \\ \hline
\end{tabular}
\end{table}
The DWBA results obtained with the complex folded OP and inelastic folded FF are
compared with the measured cross section for inelastic \ac scattering to the
Hoyle state in Figs.~\ref{f3} and \ref{f4}. We found that the calculated DWBA
cross sections systematically overestimate data at all energies if the inelastic
folded FF is obtained with the (\emph{unrenormalized}) RGM transition density
\cite{Kam81}. Given the accurate choice of the density dependent $NN$
interaction and nuclear densities, this discrepancy is clearly not associated
with usual uncertainties of the folding model analysis of \aA scattering
\cite{Sat97}. Since the RGM transition density has been proven to reproduce
nicely the $(e,e')$ data \cite{Kam81,Fun06}, the folding + DWBA results shown in
Figs.~\ref{f3} and \ref{f4} might indicate a ``suppression" of the monopole $E0$
strength occurred in the inelastic \ac scattering. To further probe this effect,
we have used a realistic Fermi distribution for the g.s. density of $^{12}$C
\cite{Kho01} to generate the BM transition density (\ref{e1}) for the DWBA
analysis of inelastic scattering to the Hoyle state. We have fixed the monopole
deformation parameter $\beta$ so that (\ref{e1}) gives exactly the same $E0$
transition strength as that given by the RGM transition density and, hence,
exhausts 22.8\% of the isoscalar monopole EWSR as discussed above. As can be
seen from Figs.~\ref{f3} and \ref{f4}, the DWBA cross sections given by the
inelastic FF folded with the BM transition density for the Hoyle state are very
close to those given by the RGM transition density.
To match the calculated DWBA cross section to the data points, we need to scale
both transition densities by a factor of 0.55 which reduces the sum rule
strength of the Hoyle state to $S_0\approx 6.9\%$ of the EWSR. Such a small sum
rule strength is significantly below the empirical range of 15 - 20\% of the
EWSR discussed above. Given an accurate energy dependence of the CDJLM
interaction, with its density dependence carefully calibrated against the BHF
results at each energy, the RGM and BM transition densities scaled by the same
factor of 0.55 give reasonable DWBA descriptions of the data at four considered
energies. The sum rule strength of the Hoyle state found in our DWBA analysis
is comparable to that ($7.6\pm 0.9\%$) found in the folding + DWBA analysis of
the inelastic \ac scattering at 240 MeV \cite{John03}, where a Gaussian has
been used for the $\alpha N$ interaction in the folding calculation of the OP
and inelastic FF.

A widely used prescription for the DWBA calculation of inelastic hadron
scattering is to assume the same complex OP for the entrance and exit channels
\cite{Sat83} and the only difference in the elastic waves is a slight shift in
kinematics caused by a lower energy in the exit channel. For such a subtle
nuclear state like the Hoyle state which has unusually dilute and extended
structure \cite{Fre07}, the diagonal potential $U_{0^+_2\to 0^+_2}$ should be
quite different from $U_{0^+_1\to 0^+_1}$ which was used as the OP in both the
entrance and exit channels of the DWBA calculation presented above. Actually,
the importance of using explicitly different OP for different exit channels in
the CC description of \ac scattering at 104 and 139 MeV was pointed out earlier
by Bauhoff \cite{Bau84}. Thus, one must go beyond DWBA and do the CC analysis of
inelastic \ac scattering to the Hoyle state before making any conclusion on the
missing monopole strength. Although the $E2$ transition between $0^+_2$ and
$2^+_1$ states is quite weak and $0^+_2$ state does \emph{not} belong to a
two-phonon $(2^+_1\otimes 2^+_1)$ band, given the importance of $0^+_2\to 2^+_1$
transition in the stellar Carbon production, we have considered following
coupling scheme in our CC calculation
 \begin{equation}
  0^+_1\leftrightarrow 2^+_1\leftrightarrow 0^+_2 \leftrightarrow 0^+_1.
 \label{e7}\end{equation}
In the CC scheme (\ref{e7}), the two-step excitation ($0^+_1\to 2^+_1\to 0^+_2$)
of the Hoyle state is treated in equal footing with the direct excitation
($0^+_1\to 0^+_2$). Thus, the CC description of the inelastic \ac scattering to
$0^+_2$ state should be more realistic compared to the DWBA calculation which
takes into account the direct excitation only \cite{vOe07}. For the inputs of
the CC calculation, in addition to three inelastic FF's for transitions between
$0^+_1,\ 2^+_1$ and $0^+_2$ states, three complex OP's have been calculated
using the diagonal nuclear densities of these states given by the RGM
calculation \cite{Kam81} and CDJLM interaction (\ref{e4}). For consistency, the
$N_{\rm R(I)}$ factors have been readjusted again to obtain a good CC
description of the elastic scattering data, as shown in Fig.~\ref{f2}. While the
best-CC-fit $N_{\rm R}$ factor (see Table~\ref{t2}) remains more or less the
same, $N_{\rm I}$ factor becomes significantly smaller and closer to unity,
especially at the two low energies. This result shows explicitly how the
inelastic channels in the CC scheme (\ref{e7}) contribute to the enhanced
$N_{\rm I}$ factor found in the OM analysis of elastic scattering. With the
increasing energy, more nonelastic channels are open and the best-CC-fit $N_{\rm
I}$ factor remains around 1.2 to 1.3 at 172.5 and 240 MeV. All the CC results
are shown in the right panels of Figs.~\ref{f3} and \ref{f4}, and it can be seen
that the CC cross sections strongly overestimate the inelastic data for $0^+_2$
state, in the same way as found in the DWBA calculation, when the inelastic FF
is obtained with the original RGM transition density \cite{Kam81}. The two-step
excitation ($0^+_1\to 2^+_1\to 0^+_2$) was found to be negligible at all
considered energies and the CC cross sections shown in Figs.~\ref{f3} and
\ref{f4} are practically given by the direct (one-step) excitation of $0^+_2$
state only. We note that the original RGM density for the $2^+_1\to 0^+_2$
transition could not reproduce the observed $B(E2)$ transition probability and
we have scaled this transition density, as recommended in Ref.~\cite{Kam81}, by
a factor of 1.53 to reproduce the experimental transition rate $B(E2,2^+_1\to
0^+_2)\approx 2.6\pm 0.8\ e^2$fm$^4$ \cite{Endt79}. Despite such an enhancement,
the folded $2^+_1\to 0^+_2$ transition FF gives still a very weak contribution
to the CC cross section for the $0^+_2$ excitation. An agreement with data can
be reached only if the $E0$ transition density is reduced by about half as found
in the DWBA analysis. It is clear now that the missing monopole strength in the
inelastic \ac scattering to the Hoyle state is \emph{not} associated with
uncertainties of the DWBA or CC analyses, and there should be some physics
effect that damps the $E0$ transition strength. We further mention a recent
folding + CC calculation of inelastic \ac scattering at $E_{\rm lab}=172.5, 240$
and 386 MeV (using the same RGM nuclear densities and DDM3Y interaction) by
Takashina \cite{Tak07}, where coupling to all the low-lying excited states of
$^{12}$C as well as 3-$\alpha$ breakup channel was taken into account, but the
CC cross section for the $0^+_2$ excitation remains lying significantly higher
than the data points, in about the same manner as found in our CC analysis.

Such a ``damping" of the transition strength in inelastic \AA scattering has
also been observed in our recent folding model study \cite{Kho05} of inelastic
\oo scattering to the 2$^+_1$ and 3$^-_1$ states of $^{16}$O. Given the electric
transition strengths of these states of $^{16}$O well determined from the
$(e,e')$ data (like the Hoyle state considered here), the DWBA or CC description
of high-precision inelastic \oo scattering data, which cover a wide angular
range and 6 orders of the cross-section magnitude, was possible only if the
absorption in the exit channel is significantly increased \cite{Kho05}. Such an
enhanced absorption was also found for the exit channel of one-neutron transfer
\o17o15 reaction to the $3/2^-$ excited state of $^{15}$O \cite{Boh02}. The
enhanced absorption in the exit channel is a direct consequence of the
suppression of nuclear refraction in nonelastic channels \cite{Kho07}. Namely, a
short-lived (excited) and loosely bound cluster like $^{16}$O$^*_{2^+}$ or
$^{15}$O$^*_{3/2^-}$ has a shorter mean free path in the nuclear medium which
implies a weaker refraction or stronger absorption \cite{Jeu76,Neg81} in the
exit channel compared to the entrance channel where both nuclei are in their
(stable) ground states (for detailed discussion see
Refs.~\cite{Kho07,Kho05,Boh02}). In a similar scenario, the $0^+_2$ state of
$^{12}$C is very weakly bound and particle unstable (with a mean lifetime
$\tau\approx 10^{-16}$ s) and it is natural to expect that the OP in the \ext
exit channel is more absorbing than the OP in the entrance \ac channel. Since
the elastic $\alpha$ scattering on the Hoyle state cannot be measured, we have
adjusted the renormalization factor $N_{\rm I}$ of the imaginary OP of the \ext
channel to achieve a reasonable CC description of the inelastic data for $0^+_2$
state, while keeping $N_{\rm R}$ factor and complex strength of the transition
FF unchanged. From Figs.~\ref{f3} and \ref{f4} one can see that a very good CC
description of the inelastic $0^+_2$ data can be reached by using $N_{\rm
I}\approx 2.5 - 3.4$ for the imaginary OP in the \ext channel (see
Table~\ref{t2}). Such a strong absorptive OP predicts a very different elastic
\ext scattering cross section compared to the measured elastic \ent scattering
(see Fig.~\ref{f2}).

\begin{figure}[ht]
\centering\vspace*{3cm}%\hspace*{-2cm}
 \mbox{\epsfig{file=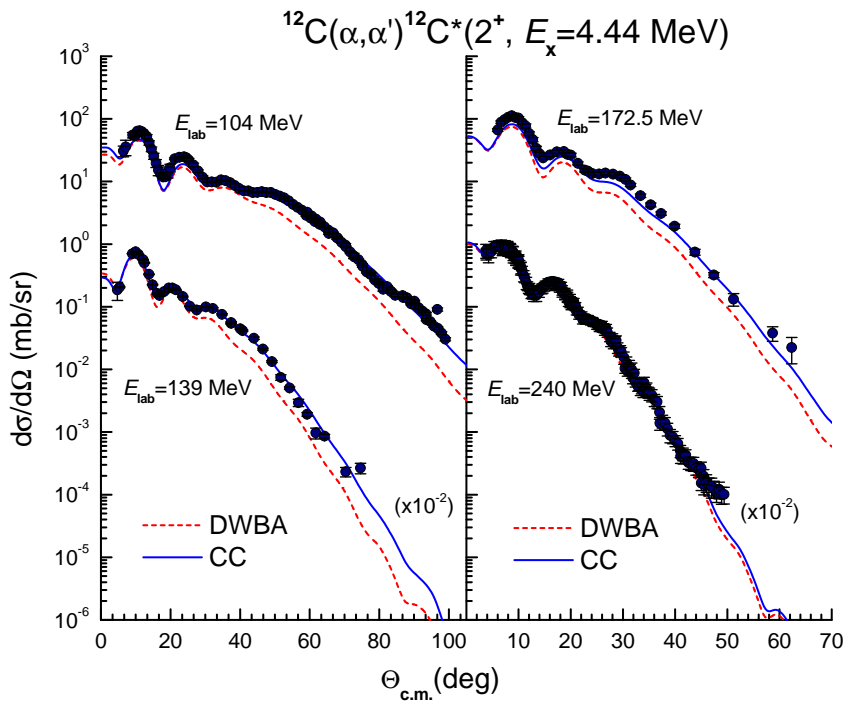,height=10cm}}\vspace*{-3.5cm}
\caption{Inelastic \ac scattering data measured at $E_{\rm lab}=104$
\cite{Hau69,Spe71}, 139 \cite{Smi73}, 172.5 \cite{Wik81} and 240 \cite{John03}
MeV for the $2^+_1$ excitation at 4.44 MeV in $^{12}$C in comparison with the
DWBA and CC results obtained with the complex folded OP and inelastic folded FF.
See more details in text.} \label{f5}
\end{figure}

To illustrate whether the enhanced absorption is also required for some other
excited states of $^{12}$C or is it a unique effect associated with fragile
structure of the Hoyle state, we have done the same folding + DWBA (CC) analyses
of inelastic \ac scattering to the $2^+_1$ state at 4.44 MeV using the RGM
density and the results are plotted in Fig.~\ref{f5}. Like the Hoyle state, the
RGM transition density of the $2^+_1$ state has also been used to successfully
reproduce the $(e,e')$ data \cite{Kam81}. The electric transition strength
predicted by the RGM transition density is $B(E2,0^+_1\to 2^+_1)=46.5\
e^2$fm$^4$ which agrees fairly with the measured $B(E2)$ transition rate of
$40\pm 4\ e^2$fm$^4$ \cite{Endt79,Ram01}. We note that $2^+_1$ state is a very
strong excitation and the measured inelastic cross section becomes even
comparable to the elastic scattering cross section at the medium angles where
the nuclear scattering dominates (see Figs.~\ref{f2} and \ref{f5}). That is the
reason why the coupling strength by the $2^+_1$ excitation is strongest in our
CC scheme (\ref{e7}). From Fig.~\ref{f5} one can see that a consistent folding
model description of the inelastic \ac scattering to the $2^+_1$ state at
$\alpha$-energy up to 172.5 MeV can be reached only in the CC calculation. If
one stays within the DWBA framework, using the same inelastic folded FF, then
the total strength of the $E2$ transition density needs to be enlarged to
reproduce the $2^+_1$ data, but this would lead to a larger discrepancy between
the calculated and experimental $B(E2,0^+_1\to 2^+_1)$ values. It can also be
seen from Table~\ref{t2} that the coupling strength by the $2^+_1$ excitation
exhausts nearly all the surface enhancement of the imaginary OP at 104 and 139
MeV, and the best-CC-fit $N_{\rm I}$ factor is around unity compared to that of
about 1.3 - 1.4 as given by the OM analysis. Finally, the measured inelastic \ac
scattering cross section for the $2^+_1$ state is nicely reproduced by the CC
calculation at all energies without the need to enhance the imaginary OP in the
\2ext channel, i.e., CC$_{\rm en}$ set of the $N_{\rm R(I)}$ factors in
Table~\ref{t2} was also used for the exit \2ext channel. This indicates that the
enhanced absorption found above for the \ext channel seems to be associated
particularly with the fragile structure and short lifetime of the Hoyle state.
We recall that the $2^+_1$ state of $^{12}$C is more robust and particle stable,
and it decays via $\gamma$-emission only. With a mean lifetime $\tau\approx
6\times 10^{-14}$ s \cite{Ram01} the $2^+_1$ state lives about 600 times longer
than the Hoyle state and 10 times longer than the $2^+_1$ state of $^{16}$O for
which the absorption enhancement has been found in the folding model study of
inelastic \oo scattering \cite{Kho05}. Although our folding model analysis was
done using a new complex density dependent interaction, the absorption
enhancement found for the exit \ext channel is not a feature associated with the
new interaction. In fact, such a large absorption has already been established
in the semi-microscopic CC analysis of the same \ac data by Ohkubo and
Hirabayashi \cite{Oh04}. These authors have used phenomenological Woods-Saxon
(WS) potentials of the same geometry for the imaginary OP in all exit channels
of the inelastic \ac scattering and adjusted the WS depth each case to reproduce
the inelastic data. From Table~I of Ref.~\cite{Oh04} one can find that the WS
imaginary OP for the \ext channel is nearly 4 times deeper than that for the
entrance \ent channel, while the best-fit WS depth found for the \2ext channel
is only slightly larger than that of the \ent channel. For the $3^-_1$ state of
$^{12}$C the CC calculation of Ref.~\cite{Oh04} has shown also a sizable
enhancement of absorption, in an agreement with the results of our CC analysis
using the CDJLM interaction which will be presented elsewhere. We note further
that small fraction ($S_0\approx 9\%$) of the monopole EWSR found in the DWBA
analyses of inelastic $^{3}$He,$^6$Li+$^{12}$C scattering to the Hoyle state
\cite{Leb80,Eyr87} could well be associated with the same absorption effect as
that found in the present work.

Our results of the CC description of inelastic \ac scattering to the Hoyle state
as well as those of inelastic scattering and one-neutron transfer reaction
measured with the \oo system \cite{Kho05,Boh02} show clearly that there is a
correlation between the weak binding and/or short lifetime of the \emph{excited}
target-like cluster and the \emph{absorption} in the exit channel of a
quasi-elastic scattering reaction. Although it is still difficult to establish a
systematics for this ``absorption" phenomenon, the results obtained sofar
indicate the need to have a realistic choice for the OP not only in the entrance
but also in the exit channel of the inelastic \AA scattering. The standard
method of choosing the same complex OP in both the entrance and exit channels of
\AA scattering might lead to a large uncertainty in the deduced transition
strength if one simply scales the inelastic FF to match the DWBA or CC results
to the measured angular distributions. In particular, this effect must be
accurately taken into account in the study  of the \AA scattering measured with
unstable beams where nuclei in the entrance and exit channels are very
differently bound.

\begin{figure}[ht]
\centering\vspace*{2.5cm}%\hspace*{-2cm}
\mbox{\epsfig{file=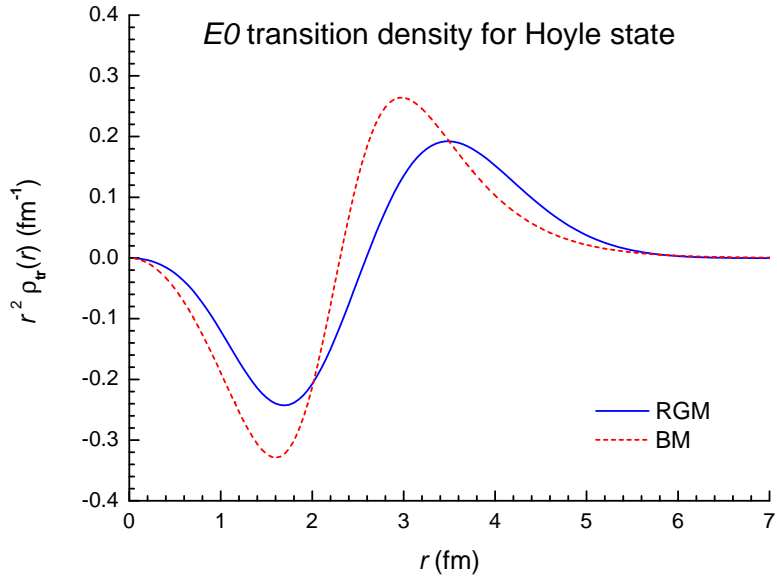,height=9cm}}\vspace*{-3cm}
 \caption{$0^+_1\to 0^+_2$ transition densities for the Hoyle state
given by the RGM calculation of Kamimura \cite{Kam81} and BM model (\ref{e1}).
The deformation parameter $\beta$ of the BM density was adjusted to reproduce
the same monopole transition strength as that given by the RGM transition
density which exhausts 22.8\% of the isoscalar monopole EWSR.}
 \label{f6}
\end{figure}
We discuss now briefly the choice for the $E0$ transition density of the Hoyle
state. It is a common belief that the breathing mode (BM) model (\ref{e1}) is
more appropriate for the transition density of a giant monopole resonance (in
medium and heavy nuclei) that exhausts a large fraction of the monopole EWSR
\cite{Ho95}. Although the BM density was scaled to reproduce the monopole
transition moment given by the RGM density \cite{Kam81} before being used in our
folding calculation, its radial dependence (see Fig.~\ref{f6}) should not
contain any feature of the $\alpha$ cluster or BEC structure of the Hoyle state.
Therefore, it is quite surprising to see that the BM transition density gives
nearly the same DWBA and CC descriptions of the inelastic $0^+_2$ data as those
given by the 3-$\alpha$ RGM transition density (see Figs.~\ref{f3} and
\ref{f4}). One can see in Fig.~\ref{f6} that the two densities have the same
asymptotic tail at $r>6$ fm while the node and two extrema of the BM density are
shifted inward to smaller radii compared to the RGM density. However, as shown
in Figs.~\ref{f3} and \ref{f4}, such a difference in the two densities gives
only a minor difference in the calculated inelastic scattering cross sections.
Consequently, the success of the BEC or RGM nuclear densities in the CC
description of inelastic \ac scattering to the $0^+_2$ state reported sofar
\cite{Hir02,Oh04,Tak06,Oh07} is an important evidence, but \emph{not} the
unambiguous proof for the $\alpha$-condensate structure of the Hoyle state. In
fact, a closer inspection of the structure models by Chernykh {\em et al.}
\cite{Che07} has revealed that the structure of the Hoyle state could be
somewhat more complicated than just a condensate of three $\alpha$ particles.

In conclusion, a realistic (complex) density dependence was introduced into the
M3Y-Paris interaction, based on the Brueckner Hartree Fock calculation of
nuclear matter, for the folding model study of the \ac scattering at $E_{\rm
lab}=104$ to 240 MeV. Given an accurate estimation of the bare \ac optical
potential, our folding model analysis has shown consistently that there should
be an enhancement of absorption in the exit \ext channel due to the short
lifetime and weakly bound structure of the Hoyle state, which accounts for the
``missing" monopole strength of the Hoyle state observed earlier in the DWBA
analysis of inelastic \ac scattering.

\section*{Acknowledgments}
The authors are grateful to Prof. W. von Oertzen, Prof. S. Ohkubo, Dr. M.
Takashina and Dr. X. Chen for their helpful communications. We also thank Prof.
M. Kamimura for providing us with the revised parametrization of the RGM
densities from Ref.~\cite{Kam81}.


\begin{thebibliography}{00}
\bibitem{Hau69} G. Hauser, R. L\"ohken, H. Rebel, G. Schatz, G.W. Schweimer,
 and J. Specht, Nucl. Phys. A 128 (1969) 81.
\bibitem{Spe71} J. Specht, G.W. Schweimer, H. Rebel, G. Schatz, R. L\"ohken,
 G. Hauser, Nucl. Phys. A 171 (1971) 65.
\bibitem{Smi73} S.M. Smith, G. Tibell, A.A. Cowley, D.A. Goldberg, H.G. Pugh,
 W. Reichart, N.S. Wall, Nucl. Phys. A 207 (1973) 273.
\bibitem{Wik81} S. Wiktor, C. Mayer-B\"oricke, A. Kiss, M. Rogge, P. Turek,
 and H. Dabrowski, Acta Phys. Pol. B 12 (1981) 491;
 A. Kiss, C. Mayer-B\"oricke, M. Rogge, P. Turek, and S. Wiktor,
 J. Phys. G 13 (1987) 1067.
\bibitem{Stre70} P. Strehl, Z. Phys. 234 (1970) 416.
\bibitem{Fre07} M. Freer, J. Phys. G 34 (2007) 789;
 M. Freer, to appear in {\it Proceedings of Intern. Symp. on Physics of
 Unstable Nuclei (ISPUN07)} (World Scientific: Singapore).
\bibitem{Ueg77} E. Uegaki, S. Okabe, Y. Abe, H. Tanaka,
 Prog. Theor. Phys. 57 (1977) 1262.
\bibitem{Kam81} M. Kamimura, Nucl. Phys. A 351 (1981) 456;
 M. Kamimura, private communication.
\bibitem{Toh01} A. Tohsaki, H. Horiuchi, P. Schuck, G. R\"opke,
 Phys. Rev. Lett. 87 (2001) 192501.
\bibitem{Fun03} Y. Funaki, A. Tohsaki, H. Horiuchi, P. Schuck, G. R\"opke,
 Phys. Rev. C 67 (2003) 051306(R).
\bibitem{Che07} M. Chernykh, H. Feldmeier, T. Neff, P. von Neumann-Cosel,
 A. Richter, Phys. Rev. Lett. 98 (2007) 032501.
\bibitem{Fun06} Y. Funaki, A. Tohsaki, H. Horiuchi, P. Schuck, G. R\"opke,
 Eur. Phys. J. A 28 (2006) 259.
\bibitem{Leb80} D. Lebrun, M. Buenerd, P. Martin, P. de Saintignon,
 G. Perrin, Phys. Lett. 97 B (1980) 358.
\bibitem{Eyr87} W. Eyrich, A. Hofmann, A. Lehmann, B. M\"uhldorfer, H.
 Schl\"osser, H. Wirth, H. J. Gils, H. Rebel, S. Zagromski,
 Phys. Rev. C 36 (1987) 416.
\bibitem{John03} B. John, Y. Tokimoto, Y.W. Lui, H.L. Clark, X. Chen,
 D.H. Youngblood, Phys. Rev. C 68 (2003) 014305.
\bibitem{Sat83} G.R. Satchler, {\it Direct Nuclear Reactions}
 (Clarendon Press: Oxford, 1983).
\bibitem{Bau84} W. Bauhoff, Phys. Lett. 139 B (1984) 223.
\bibitem{Hir02} Y. Hirabayashi, S. Ohkubo, Phys. At. Nuclei 65 (2002) 683.
\bibitem{Oh04} S. Ohkubo, Y. Hirabayashi, Phys. Rev. C 70 (2004) 041602(R).
\bibitem{Tak06} M. Takashina, Y. Sakuragi, Phys. Rev. C 74 (2006) 054606.
\bibitem{Oh07} S. Ohkubo, Y. Hirabayashi, Phys. Rev. C 75 (2007) 044609.
\bibitem{Ko82} A.M. Kobos, B.A. Brown, P.E. Hodgson, G.R. Satchler,
 A. Budzanowski, Nucl. Phys. A 384 (1982) 65.
\bibitem{Ho95} D.J. Horen, J.R. Beene, G.R. Satchler,
 Phys. Rev. C 52 (1995) 1554.
\bibitem{Kho97} D.T. Khoa, G.R. Satchler, W. von Oertzen,
 Phys. Rev. C 56 (1997) 954.
\bibitem{Kho00} D.T. Khoa, G.R. Satchler, Nucl. Phys. A 668 (2000) 3.
\bibitem{Sat97} G.R. Satchler, D.T. Khoa, Phys. Rev. C 55 (1997) 285.
\bibitem{Je77} J.P. Jeukenne, A. Lejeune, C. Mahaux, Phys. Rev. C 16 (1977) 80.
\bibitem{An83} N. Anantaraman, H. Toki, and G.F. Bertsch,
 Nucl. Phys. A 398 (1983) 269.
\bibitem{Kho07} D.T. Khoa, W. von Oertzen, H.G. Bohlen, S. Ohkubo,
 J. Phys. G 34 (2007) R111.
\bibitem{Kho01} D.T. Khoa, Phys. Rev. C 63 (2001) 034007.
%\bibitem{Si76} I. Sick, J.S. McCarthy, R.R. Whitney, Phys. Lett. B 64 (1976) 33.
\bibitem{Sri84} D.K. Srivastava, H. Rebel, J. Phys. G 10 (1984) L127.
\bibitem{Raynal} J. Raynal, Computing as a Language of Physics (IAEA, Vienna,
 1972) p.75; J. Raynal, coupled-channel code ECIS97 (unpublished).
\bibitem{Ba89} C.J. Batty, E. Friedman, H.J. Gils, H. Rebel,
 Adv. Nucl. Phys. 19 (1989) 1.
\bibitem{Fe92} H. Feshbach, {\it Theoretical Nuclear Physics} Volume II
(Wiley-Interscience: New York, 1992).
\bibitem{vOe07} W. von Oertzen, private communication.
\bibitem{Endt79} P.M. Endt, At. Data and Nucl. Data Tables 23 (1979) 3;
 F. Ajenberg-Selove, Nucl. Phys. A 506 (1990) 1.
\bibitem{Tak07} M. Takashina, talk presented at DREB07 Workshop, June 2007.
\bibitem{Kho05} D.T. Khoa, H.G. Bohlen, W. von Oertzen, G. Bartnitzky,
A. Blazevic, F. Nuoffer, B. Gebauer, W. Mittig, P. Roussel-Chomaz,
 Nucl. Phys. A 759 (2005) 3.
\bibitem{Boh02} H.G. Bohlen, D.T. Khoa, W. von Oertzen, B. Gebauer, F. Nuoffer,
 G. Bartnitzky, A. Blazevic, W. Mittig, P. Roussel-Chomaz,
 Nucl. Phys. A 703 (2002) 573.
\bibitem{Jeu76} J.P. Jeukenne, A. Lejeune, C. Mahaux, Phys. Rep. 25 (1976)
 83; Phys. Rev. C 16 (1977) 80.
\bibitem{Neg81} J.W. Negele, K. Yazaki, Phys. Rev. Lett. 47 (1981) 71.
\bibitem{Ram01} S. Raman, C.W. Nestor Jr., P. Tikkanen,
 At. Data and Nucl. Data Tables 78 (2001) 1.
\end{thebibliography}
\end{document}